\documentstyle[preprint,aps,psfig]{revtex}
\begin{document}
\draft

\title{Nonequilibrium molecular dynamics simulation
of rapid directional solidification.}
\author{Franck Celestini$^*$ and Jean-Marc Debierre}
\address {Laboratoire Mat\'eriaux et Micro\'electronique de Provence$\dag$,
Universit\'e d'Aix-Marseille III,
Facult\'e des Sciences et Techniques de Saint-J\'er\^ome, Case 151,
13397 Marseille Cedex 20, FRANCE\\
}

\maketitle

\vskip1pc

\begin{abstract}
{We present the results of non-equilibrium molecular dynamics
simulations for the growth of a solid binary alloy from its liquid
phase. The regime of high pulling velocities, $V$, for which
there is a progressive transition from solute segregation
to solute trapping, is considered.
In the segregation regime, we recover the exponential
form of the concentration profile within the liquid phase.
Solute trapping is shown
to settle in progressively as $V$ is increased and our results are in good
agreement with the theoretical predictions
of Aziz [J. Appl. Phys. {\bf 53}, 1158 (1981)]. In addition,
the fluid advection velocity is shown to remain
directly proportional to $V$, even at the highest velocities
considered here ($V\simeq10$ms$^{-1}$).}
\end{abstract}


\vskip2pc

\section{Introduction}

Directional solidification (DS) of binary alloys is a reference
experimental method to conduct carefully controlled tests of industrial casting.
Besides their practical interest, DS experiments also bring insight
into the fundamental study of basic instability morphologies of solid-liquid
interfaces, such as cells or dendrites \cite{Bernard_book}.
By the combination of theoretical and numerical methods, much progress has been
now achieved toward solving this difficult physical problem.
However, most
of the theoretical effort has concentrated so far on continuous descriptions
of the underlying phenomena \cite{Klaus}. In contrast, attempts to
attack the problem at the atomistic level remain very few.
To this extent, the present study may be considered as a first step
to bridge the gap between both points of view.

Because of the micrometer size of the growth structures,
a direct and quantitative atomistic simulation
is still not at hand on this scale.
Nevertheless, in the limit of large pulling
velocities, a microscopic technique like Molecular Dynamics (MD)
can still be used to follow atomistic
phenomena occuring close to the solid-liquid interface.
On one hand, MD was used to simulate laser-pulsed melting, for which the
velocity is not controlled, but rather governed by heat diffusion.
\cite{Abraham,Chokappa,Richardson}. On the other hand, a first
MD simulation of directed growth has been recently reported
by Coura and al. \cite{Coura}.
They considered the growth of a solid from a fluid phase
with a density ten times smaller, a case which is probably
more relevant to deposition from a vapor.

In this paper, we present non-equilibrium molecular dynamic simulations of
directional solidification in two dimensions. We restrict ourselves to the
case of
rapid solidification with a large temperature gradient, for which the interface
remains plane on the atomistic scale, so that microstructures will not be
considered here.
After the description of the simulation details in section II, we present
in section III
the results obtained for the segregation profiles,
the segregation coefficient and the advection velocity. In the last section, we
finally discuss how this nonequilibrium simulation can be extended in the
future to study different microscopic mechanisms involved in DS.

\section {Simulation details}

Two atoms $i$ and $j$ separated by a distance $r$ interact via the well-known
Lennard-Jones potential :
\begin{equation}
u_{ij}(r)=4\epsilon_{ij}[(\frac{\sigma_{ij}}{r})^{12}-(\frac{\sigma_{ij}}{r})^{6
}]
\end{equation}
The interaction energies between pairs of
solvent-solvent, solute-solute and solvent-solute atoms
are respectively denoted by $u_{11}$, $u_{22}$ and $u_{12}$.
The solvent potential parameters
are chosen in order to describe Argon properties ($\epsilon_{11}=120$K,
$\sigma_{11}=3.405 \rm\AA$). For the solute,
we take $\epsilon_{22}=0.5 \epsilon_{11}$ and $\sigma_{22}=\sigma_{11}$.
The cross-species parameters  are fixed using the Lorentz-Berthelot rules
\cite{Lorentz} :
\begin{equation}
\epsilon_{12}=\sqrt{\epsilon_{11}\epsilon_{22}}
\end{equation}
and
\begin{equation}
\sigma_{12}=(\sigma_{11}+\sigma_{22})/2.
\end{equation}
All the interactions are truncated at a cut-off radius $r_c=2.5\sigma_{11}$ and
the equations of motion are integrated using the Beeman algorithm
\cite{Beeman} with a time step $\delta t=0.01$ps.
The particles coordinates are defined in a reference frame moving at the pulling
velocity $V$ in the $x$ direction and periodic boundary conditions (PBC) are
applied in the $x$ and $y$ directions. After integration of the dynamical
equations
over a time $\delta t$, pulling is implemented by adding
an increment $-V\delta t$ to the $x$ coordinate of each atom.
In the reference frame, the solid-liquid interface is thus immobile
when the stationnary state is reached.

To simulate heat transport from the furnace to the system, four regions
of fixed temperature are used (Fig. \ref{temp_prof}). Regions I, II, III and IV
are centered at fixed positions, $x_I$, $x_{II}$, $x_{III}$ and $x_{IV}$
and have a width
of $20 \rm\AA$ for regions I and IV and $10 \rm\AA$ for regions II and III.
In each region the temperature is kept constant by using a classical velocity
rescaling. To maintain the solidification front between region II and III,
we impose
$T_I=T_{II}<T_m$ and $T_{III}=T_{IV}>T_m$, $T_m$ being the melting temperature
of the alloy.

In Fig. \ref{temp_prof} we represent the temperature gradient obtained after
equilibration in the simulation box.
The large difference between $T_{II}$ and $T_{III}$ and hence the large
gradient
permits to localize the interface easily. With such a high gradient,
instabilities
cannot develop, which ensures the stability of planar interfaces.
For the same reason we also take a small width $L_y$ for the simulation box in
order to reduce the natural roughness of the interface. More realistic systems,
as compared to experiments, would correspond to $x_I=x_{II}$ and
$x_{III}=x_{IV}$,
together with a larger value of $L_y$. Because of the PBC in the $x$ direction,
we simultaneously have a solidification and a melting front, as in a melting
zone experiment. We concentrate here on the
solidification front and we use a wide liquid zone
($x_{III}<x<x_{IV}$) to allow solute diffusion.
Fixing the temperature in four regions instead of two
reduces considerably the rescaling of velocities within the liquid region.
This is helpful to suppress pertubations and artefacts during the computation
of microscopic quantities. The density difference between the solid and the
liquid ($\simeq 20\%$) is sufficient to induce advection of the liquid
towards the front. A part of the associated momentum is then transmitted to the
solid layer which in turn acquires a translationnal motion in the $x$ direction.
To avoid this finite-size effect, we rescale to zero the
mean velocity within the deeper part of the solid (region I).

Our system contains about $2000$ atoms, 10 percent of which are
solute atoms. Its size, $L_x=400 \rm\AA$ in length and $L_y=60 \rm\AA$ in width,
is relatively modest as compared to the size of systems currently
used in MD. The reason is that
the computational effort is here essentially spent in the time length of
the simulation.
To obtain good statistics, the simulation has to be long enough to allow
each atom to perform several solidification-melting cycles. Since ten
cycles require a time of $10 L_x/V$, for the
slowest velocity studied here ($V=10$cms$^{-1}$), this represents a simulation
time of $4\times 10^{-4}$s ($4\times 10^8$ MD steps). To increase the
performance of
our code we then adopt the 'cell lists'  method \cite{Frenkel} in which the box
is divided into cells with a size slighty larger than the cutoff radius $r_c$.

\section{Results}
\subsection{Concentration profiles}
The partition ratio, or segregation coefficient, is defined as
\begin{equation}
k=c_s^i/c_l^i,
\end{equation}
$c_l^i$ and $c_s^i$ being the concentrations at the interface, respectively
in the liquid and the solid. With our
choice of the potential parameters ($\epsilon_{12}<\epsilon_{11}$), $k$ is
expected to be less than one \cite{Hitchcock}, so that solute
accumulates in the liquid near the front. We first illustrate how the simulation
reproduces solute rejection. The pulling
velocity is fixed to $V=1$ms$^{-1}$ and
the solute atoms are initially placed at random in the simulation box.
Fig. \ref{deux_photos}a is a snapshot of the system in its initial
configuration.
After a time $\Delta t=5$ns, which corresponds to a
spatial translation of the furnace of approximatly $L_x/8$, the
second snapshot (Fig \ref{deux_photos}b) shows
that almost all the solvent atoms initially close to the front
are incorporated into the solid. Conversely,
because of their poorer solubility in the solid phase, a majority of
the solute atoms remain in the liquid phase. As compared to
a solvent atom, it takes a longer time for a solute atom to cross
the interface and hence the solute concentration is higher at the interface, as
expected here. To quantify this segregation, we compute the profile
concentration by averaging over $10^4$ uncorrelated spatial configurations
(Fig. \ref{prof}). The diffusion equation for the solute in the moving frame
reads
\begin{equation}
 \frac{\partial c}{\partial t}= D \frac{\partial^2 c}{\partial x^2} + V
\frac{\partial
c}{\partial x}.
\end{equation}
In the stationary regime, it gives the well known theoretical profile
for the solute concentration in the liquid,
\begin{equation}
 c(x)=c_s^i+(c_l^i-c_s^i)\exp(-x/\l_s)
\end{equation}
where $\l_s=D/V$ is the diffusion length and the origin of the $x$
axis is placed at the front. A good agreement is found between
this expression and our simulations,
the best fit giving a diffusion coefficient $D_{fit}=0.3 \rm\AA^2 ps^{-1}$. To
verify this value, we independently compute the diffusion coefficient
profile $D(x)$,
with the help of Einstein's relation between the mean
square displacement of the atoms and $D$,
\begin{equation}
<\vert{\bf r}(0)-{\bf r}(t)\vert^2> = 4 D t.
\end{equation}
In Fig. \ref{dif}, we see that the diffusion coefficient increases abruptly
between zero
in the solid phase to roughly $0.4 \rm\AA^2 ps^{-1}$ in the liquid near
the interface. This variation of $D$ takes place over a distance of $30
\rm\AA$ that can be considered
as a good aproximation of the interface width. In Eq. (5), $D$ was assumed
constant.
We can check that the fitted value $D_{fit}$ approximately corresponds
to the average of $D(x)$ within the liquid
region near the  interface. The same analysis is repeated for different pulling
velocities between $V=0.1$ms$^{-1}$ and $V=5$ms$^{-1}$. For each velocity,
we compute the
concentration profile and extract $D_{fit}$ from the diffusion length.
In Fig. \ref{DfctV}, $D_{fit}$ is plotted as a function of the pulling velocity.
As discussed below, the segregation coefficient
tends to unity for the largest $V$ and the calculation of $D_{fit}$ is
thus restricted to velocities $V<3$ms$^{-1}$. A reasonable
agreement is found in each case between $D_{fit}$ and the average
value of the $D(x)$ profile.

In Fig \ref{2prof}, we plot the two concentration profiles obtained
for $V=1$ms$^{-1}$ and $V=0.5$ms$^{-1}$. As expected, the
smaller the velocity, the larger $\l_s$. If the agreement
with the exponential law is
good for the largest velocity, we notice that it
is less satisfactory for the smaller one. This can be understood
by a second look at Fig. \ref{dif}. We observe that the thermal gradient induces
an increase of the diffusion coefficient, as we go deeper in the liquid.
As a consequence, we also have an increase of the diffusion length,
which explains the disagreement observed for the largest $x$ values.

\subsection{Segregation coefficient}
Now that we verified the ability of our method to reproduce both the segregation
at the interface and the dependence of the diffusion length on
the pulling velocity, we examine the behavior of the segregation coefficient $k$
as a function of $V$. We see in Fig. \ref{2prof} that
the liquid concentration at the interface, $c_l^i$, is larger for a lower
value of $V$. This can be easily explained by the fact that, at
large pulling velocities, the solute does not have enough time to fully
diffuse over the interface width before it is incorporated into the solid.
As a consequence, some solute is trapped in the solid and segregation at
the interface
is lowered. Several models have been proposed to describe
solute-trapping \cite{Aziz,Wood,Baker}. The usual qualitative criterion
for segregation is to compare two different
characteristic times. One, $t_V=\delta_i/V$, corresponds to the growth of the
solid over a distance comparable to the interface width, $\delta_i$.
The second, $t_D=\delta_i^2/D$, is the time needed for a solute
atom to diffuse over the same distance $\delta_i$.
The Aziz model \cite{Aziz} predicts then
\begin{equation}
 k=\frac{k_e+\beta}{1+\beta}
\end{equation}
for the segregation coefficient. Here $k_e$ is the segregation coefficient
at equilibrium (i.e., when $V=0$) and
\begin{equation}
\beta=t_D/t_V=\delta_i/\l_s
\end{equation}
is the ratio of the two characteristic times (or lengths) defined above.
Two remarks arise at this point. The first one concerns the neighborhood
of the absolute stability threshold, $V=V_a$, where
the diffusion length becomes comparable to the
capillary length, $l_s\simeq d_0$. When $d_0\gg \delta_i$, one also has
$\beta\ll1$, so that $k\simeq k_e$. In this case, the absolute stability
threshold is well described by the usual linear stability analysis (see
\cite{Klaus}).
This is not true anymore, when $d_0$ becomes comparable to $\delta_i$, as
it is very likely the case for some reference alloys \cite{Succino}.
The second remark is that the assumption of continuous growth,
on which Eq. (8) relies, is actually verified by the rough solid-liquid
interfaces produced with our Lennard-Jones potential.

In Fig \ref{k} we plot our estimates for $k$ as a function of the pulling
velocity. The line is a best fit to the Aziz model.
The parameter $\beta$ is estimated by using $\delta_i=30\rm\AA$, the
characteristic width of the interface extracted from the $D(x)$ profile
in Fig.\ref{dif} and the value of the diffusion
coefficient, $D=0.2\rm\AA^2$ps$^{-1}$ is estimated at the center of this
interface region. The agreement between
the simulated values and the model is rather good for a fit with a single
free parameter. We obtain the value $k_e=0.503$ which is difficult
to confirm because it is not possible to simulate segregation at much
lower velocities with current computing power.
Simulations at a fixed pulling velocity thus provide an
indirect way of determining $k_e$. The fact that the interface is moving permits
all atoms, and especially the solute ones, to be part of the fluid phase
and then to speed up relaxation to the equilibrium situation. For an
immobile interface, one would have to wait for the very slow diffusion of
solute atoms within the solid phase. The solid-liquid equilibrium for
3D binary Lennard-Jones mixtures has been  determined by
Monte-Carlo simulations \cite{Hitchcock,Cottin}. It would be interesting
to have similar 2D results available to test our indirect evaluation of $k_e$.

\subsection{Advection velocity}
We finally study the advection of the fluid near the solidification front.
The Lennard-Jones mixture considered in this work produces a substantial
density difference between the solid and the liquid. This is illustrated
in Fig. \ref{dens}, where we plot the density profile along the $x$ direction
perpendicular to the interface. From these data, we estimate both
densities at the interface, $\rho_s^i$ and $\rho_l^i$.
We find that they do not depend on the pulling velocity and that the
normalized density difference is also a constant,
$(\rho_s^i-\rho_l^i)/\rho_l^i\simeq0.24$.
This rather high value for a Lennard-Jones potential is due to the large
temperature gradient imposed at the interface. As a consequence,
the solid cannot grow unless the liquid is advected toward the front with a
velocity $V_{ad}$. In the laboratory frame, the solid is at rest, while
the liquid has an advection velocity $V_{ad}$ all along the $x$ axix (Fig.
\ref{Vprof}).
Mass conservation at the interface imposes that $V_{ad}^i$ varies with $V$ as
\begin{equation}
 V_{ad}^i=\frac{\rho_s^i-\rho_l^i}{\rho_l^i} V.
\end{equation}
A linear variation is effectively recovered in our simulations (Fig. \ref{Vad}),
with a slope comparing well to the estimate extracted from the density profile
(see Fig. \ref{dens}).
It is interesting to note that there is no deviation from the linear law,
even at the highest velocities studied in this work. On one side, one could have
expected that the onset of solute trapping would modify the densities.
This is not the case here, probably because the solvent and solute atoms
have the same atomic radius, $\sigma_{22}=\sigma_{11}$.
On the other side, a deviation would also appear as $V$ is increased,
if the solid was created with more and more defects.
We would then have a decrease of $V_{ad}^i$ because the density of the
solid would also decrease. This type of deviation is not observed any more
in our simulations.
We finally see in Fig. \ref{Vprof} that the
advection velocity increases in magnitude with $x$.
As we go deeper in the liquid, the density
progressively decreases in the thermal gradient and
an extra advection velocity is thus necessary to compensate the density
drop between two adjacent slices of alloy.

\section{Conclusion}

With the help of MD simulations, we directly observed
and analyzed quantitatively two important
phenomena occuring at the atomistic level in rapid DS.
The first one is the cross-over between the
regime of solute segregation and the regime of solute trapping,
which sets in as the pulling velocity, $V$, is increased.
The good quantitative
agreement of our estimates for the segregation coefficient,
$k(V)$,
with existing theories \cite{Aziz,Wood,Baker} confirmed
that MD is a valuable simulation tool for DS. The second point is the
advection of the liquid phase towards the solidification
front. The advection velocity at the interface
was shown to remain proportional to $V$,
in the whole range of values considered here.
This effect is more likely masked by stronger
phenoma like convection in the usual experimental
set-ups. It may however become relevant in
micro-gravity or for thin sample DS.

The use of Lennard-Jones potentials is rather
restrictive to compare the present results with experiments.
In the future, potentials based on empirical descriptions,
like the embedded atom method \cite{Erco}, the glue model \cite{Daw},
or the effective medium theory \cite{Jacob} will be used
to study technical materials like metals.
This will allow us
to explore important issues like anisotropy and
facetting, or the role of defects and constraints.
In parallel, it will be necessary to simulate more
extended systems, in order to explore the
front instabilities from an atomistic point of view.
In this last category, we can cite the oscillations of the
solidification front just above the absolute stability
velocity \cite{MKK,Karma}.

\acknowledgments

It is a pleasure for us to thank B. Billia, J. Jamgotchian, and
H. Nguyen Thi for valuable
discussions, as well as K. Kassner for a critical reading of the manuscript.

\begin{figure}
\caption{Temperature profile along the $x$ axis. The
different regions of the simulation box in which the temperature is kept fixed
are bounded by the dotted lines. The melting temperature is $T_m\simeq 40K$.}
\label{temp_prof}
\end{figure}

\begin{figure}
\caption{ Snapshots of the system at two different times. Large and small
radius
respectively
correspond to solvent and solute atoms. a) At $t=0$, solute atoms are randomly
placed in the simulation box. One slice of liquid in contact with the interface
is marked (dark grey). b) At time $t=5$ns, almost all the solvent atoms have
solidified while a majority of the solute atoms remained liquid.}
\label{deux_photos}
\end{figure}

\begin{figure}
\caption{Concentration profile $c(x)$ along the pulling direction,
perpendicular to the
interface.}
\label{prof}
\end{figure}

\begin{figure}
\caption{Diffusion profile $D(x)$ along the direction perpendicular to the
interface. The horizontal dotted line is the average
diffusion coefficient obtained by fitting the concentration
profile of Fig. 3 to an exponential law.}
\label{dif}
\end{figure}

\begin{figure}
\caption{The average diffusion coefficient as a function of the pulling
velocity.}
\label{DfctV}
\end{figure}

\begin{figure}
\caption{Solute concentration profiles obtained for $V=1$ms$^{-1}$
(circles) and
$V=0.5$ms$^{-1}$ (diamonds).
The continuous lines are best fits to Eq. (6).}
\label{2prof}
\end{figure}

\begin{figure}
\caption{Variations of the segration coefficient with the pulling velocity. The
solid line is the best fit to the Aziz equation.}
\label{k}
\end{figure}

\begin{figure}
\caption{Density profile $\rho(x)$ along the direction perpendicular to the
interface. The upper and
lower arrows indicate respectively $\rho_s^i$ and $\rho_l^i$ at the
solid-liquid interface.}
\label{dens}
\end{figure}

\begin{figure}
\caption{Velocity profiles, $V_x(x)$, along the direction perpendicular to
the interface. Open and
filled circles respectively correspond to $V=3$ms$^{-1}$ and $V=5$ms$^{-1}$}
\label{Vprof}
\end{figure}

\begin{figure}
\caption{Interface advection velocity $V_{ad}^i$, plotted as a function of
the pulling
velocity $V$. The solid line corresponds to Eq. (10).}
\label{Vad}
\end{figure}



\begin{references}
\bibitem{Succino} For the succinonitrile-10\% aceton alloy, the
capillary length is about 3 $\AA$.
\end{references}
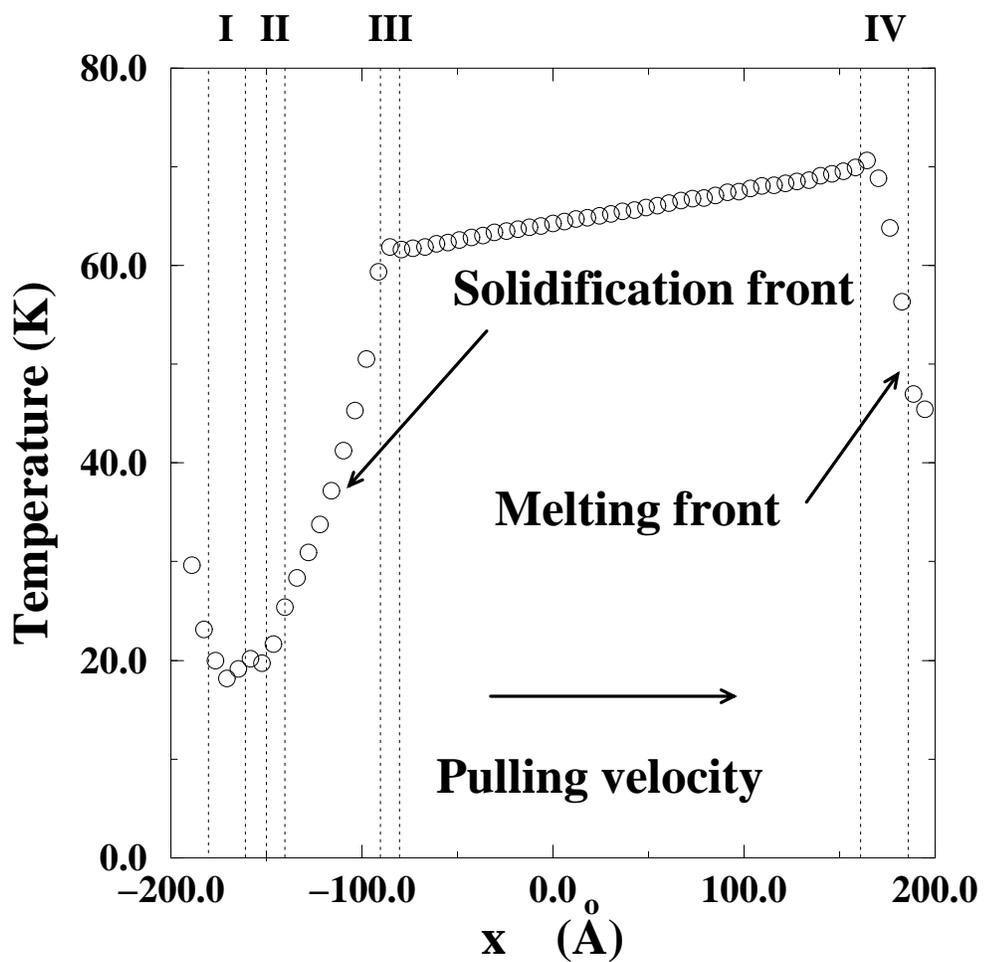
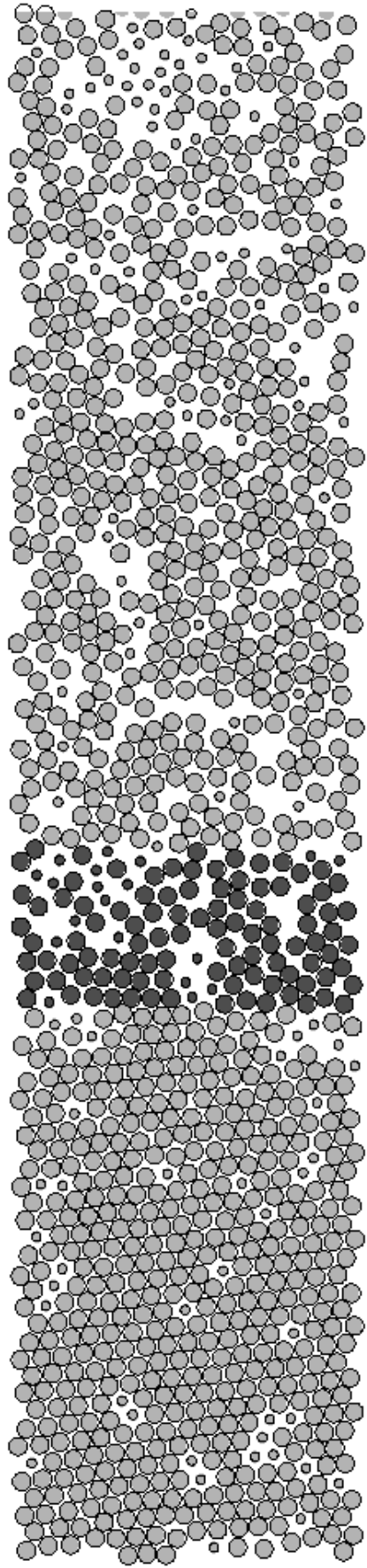
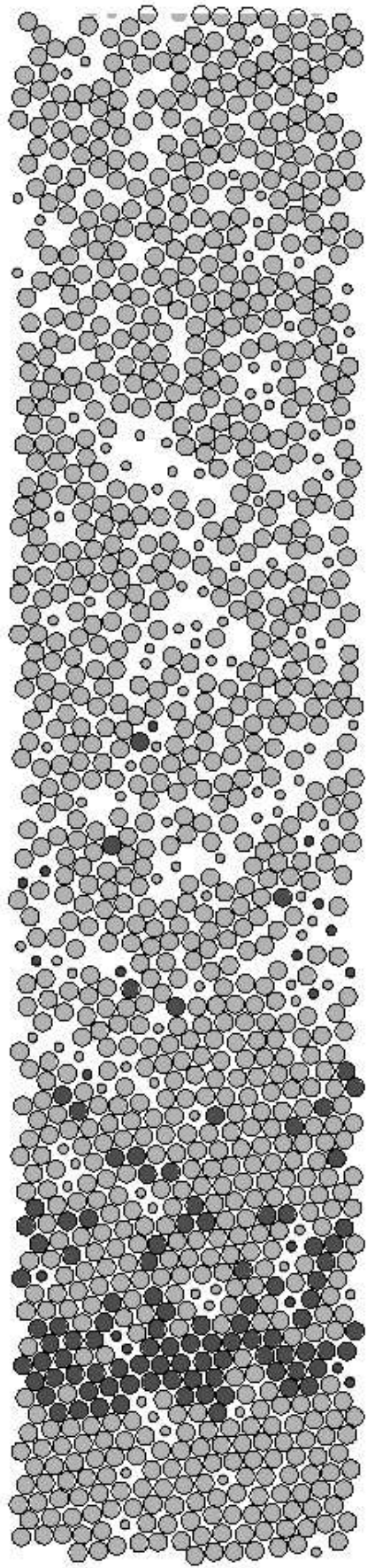
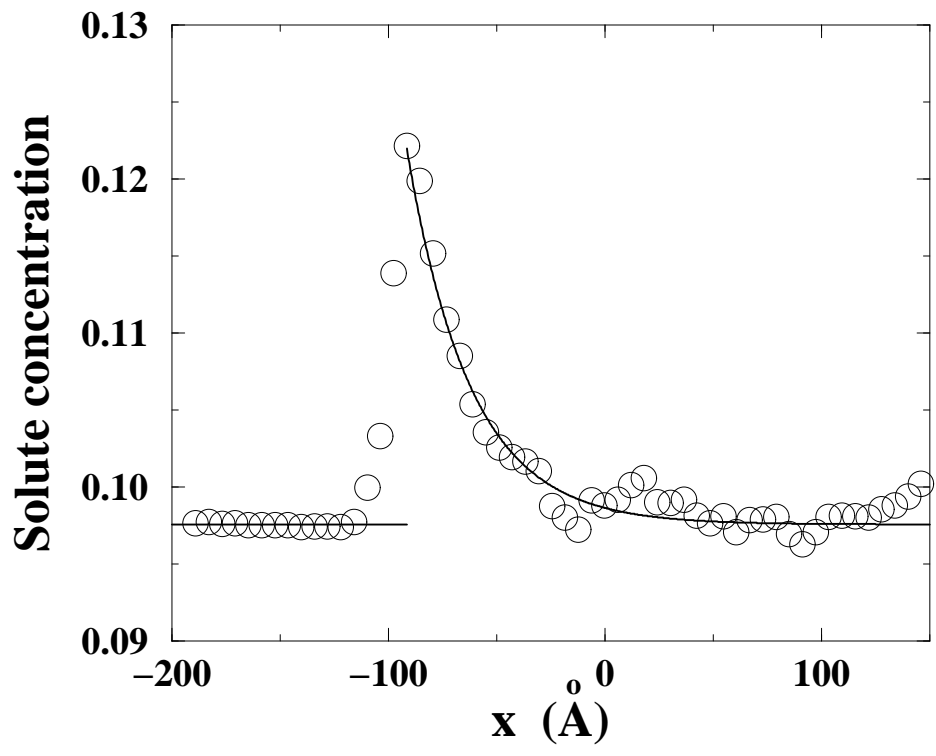
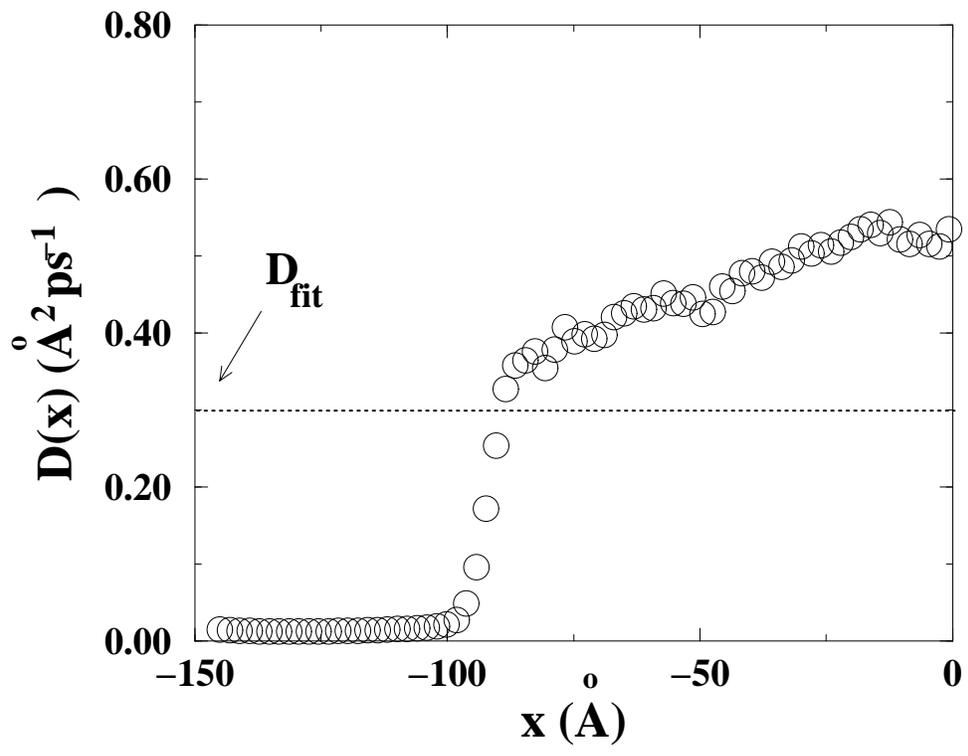
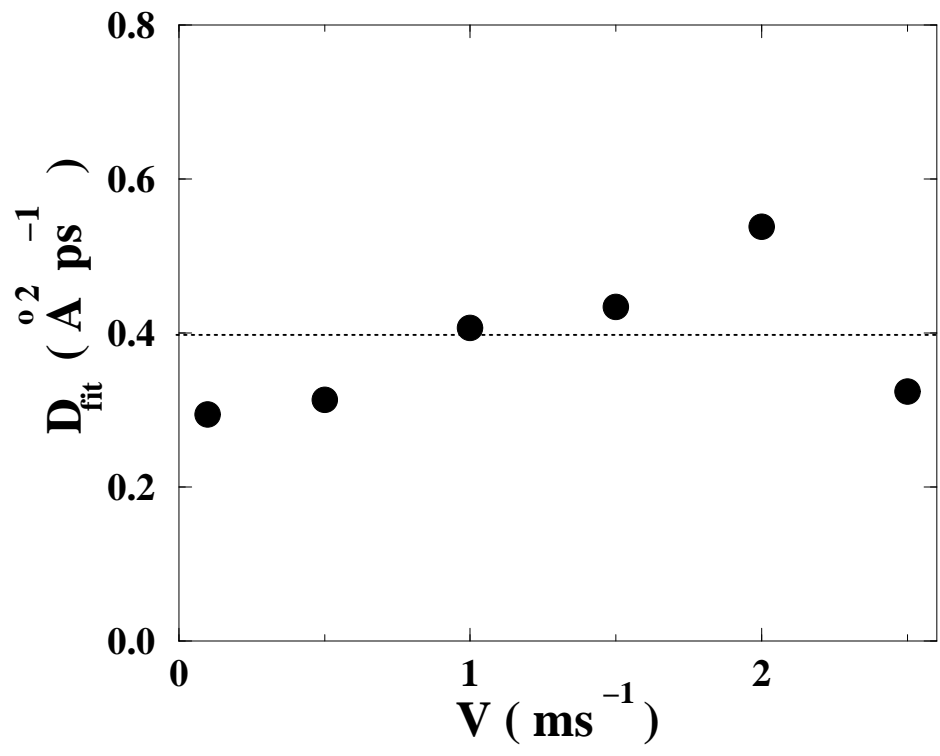
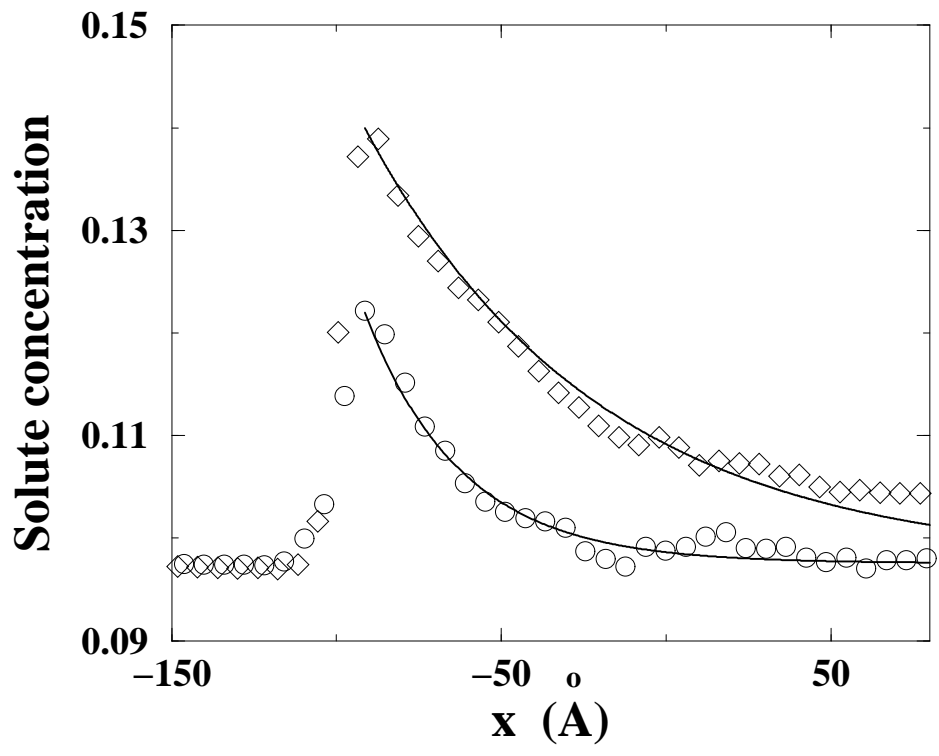
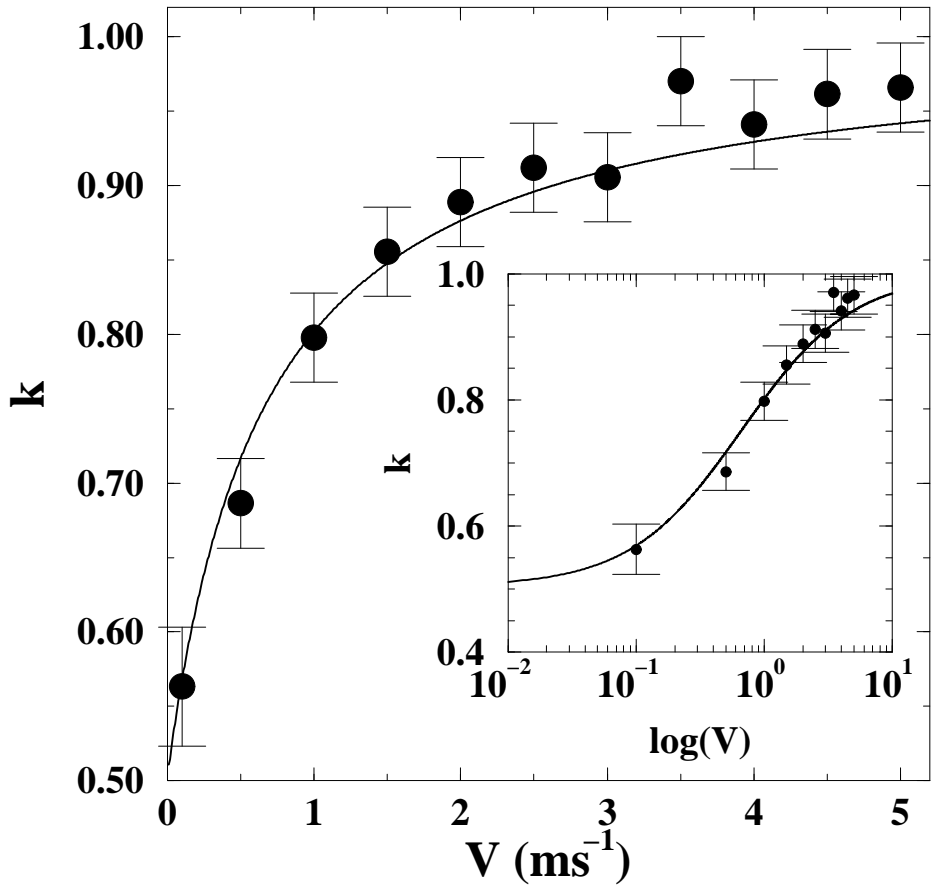
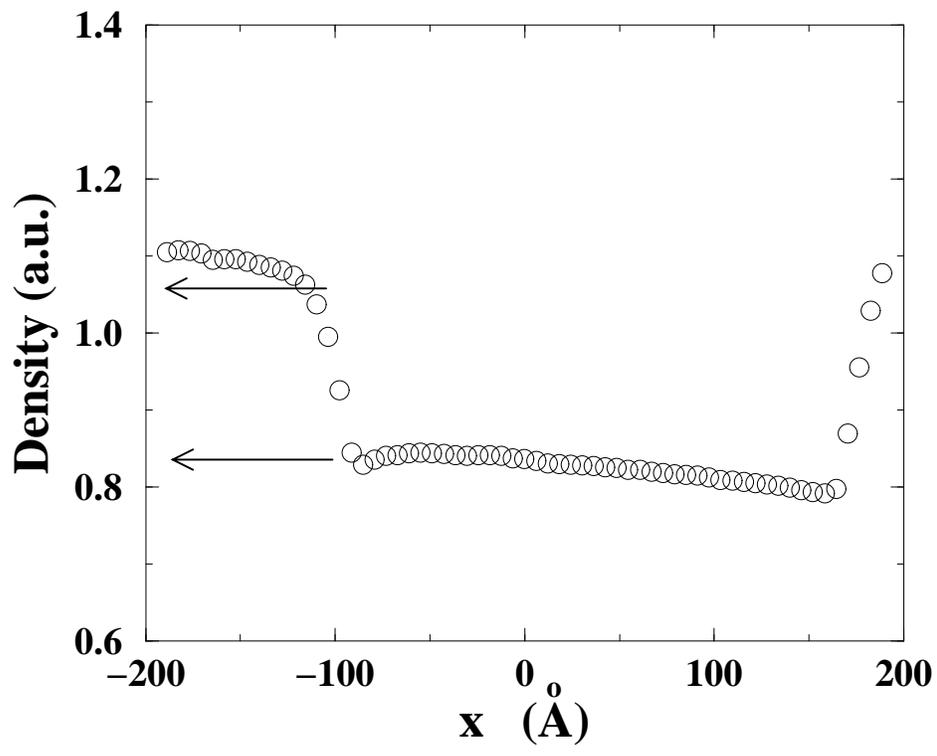
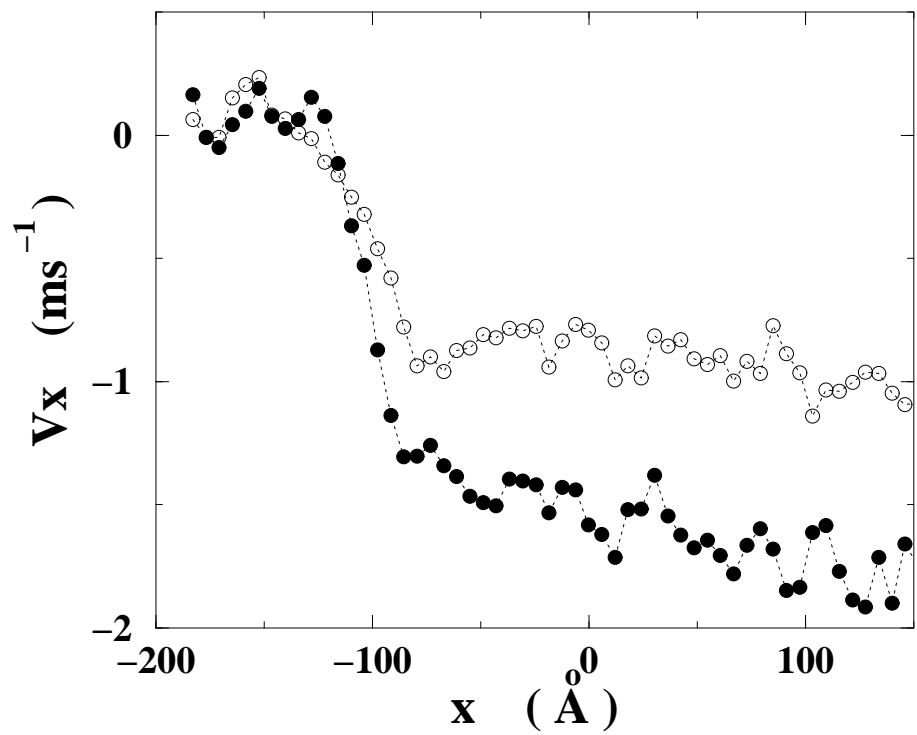
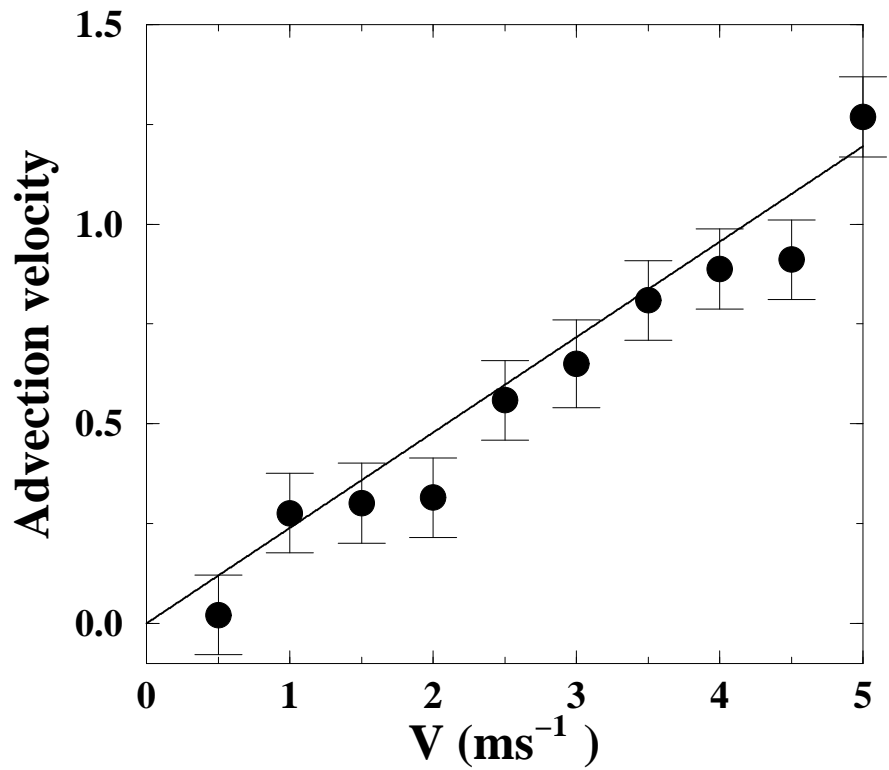
\end{document}